\documentclass[12pt]{iopart}
\usepackage{bm}
\usepackage{harvard}
\usepackage{amssymb}
\usepackage
{hyperref}

\begin{document}
\title[Another method to solve Dirac's one-electron equation numerically]
{Another method to solve Dirac's one-electron equation
numerically}

\author{K V Koshelev}

\address{ Petersburg Nuclear Physics Institute,
Gatchina 188300, Russia}

\ead{kirvkosh@gmail.com, koshelev@landau.phys.spbu.ru}

\date{\today }

\begin{abstract}
One more mode developed to get eigen energies and states for the
one-electron Dirac's equation with spherically symmetric bound
potential. For the particular case of the Coulomb potential it was
shown that the method is free of so called spurious states. The
procedure could be adapted to receive highly exited states with
great precision.
\end{abstract}


\section{Introduction}
\label{intro}
It's difficult to overestimate the importance for the relativistic
calculations numerical approaches to get full spectra of Dirac's
equation. One of the most successful is so called B-spline
approach \cite{Boor}, \cite{John}. While numerical implementation
of the problem so called spurious states can arise. Those states
need special treatment \cite{Shab}. All said above prompts to
explore new methods to solve Dirac's equation numerically. Namely
in this paper we'll investigate the advantages that squared
Hamiltonian is able to give for reaching the goal. Throughout in
the paper atomic units system ($\hbar=e=m$) is utilized.

\section{Squared Dirac's Hamiltonian}
\label{basic}
Let's take the squared one-electron Dirac's Hamiltonian in the
form:

\begin{equation}\label{Ham0}
H=(h-\epsilon)^2
\end{equation}

\noindent Where $h=c({ \alpha\cdot p})+V+mc^2\beta$ is a
well-known representation of the one-electron Dirac's Hamiltonian
\cite{Lab} with some bound potential $V$ and $\epsilon$ is an
arbitrary real-valued parameter. Having deployed the (\ref{Ham0})
everybody can easy receive
\begin{equation}\label{Ham1}
\begin{array}{l}
H=c^2({ \alpha\cdot p})^2+m^2c^4+V^2+mc^3\{({ \alpha\cdot p}),
\beta \}+c\{({ \alpha\cdot p}), V \}+ \\ \qquad 2mc^2\beta
V-2\epsilon h+\epsilon^2
\end{array}
\end{equation}
where $\{ a,b\}=ab+ba$ is a positive commutator of the operators
$a$ and $b$. Keeping in mind that the fourth term in the above
formulae (\ref{Ham1}) amounts to zero (due to the properties of
the Dirac's matrixes) one can finally get
\begin{equation}\label{Ham2}
\begin{array}{l}
H=-\hbar^2c^2\vartriangle+m^2c^4+V^2+2mc^2\beta V -i\hbar c\{({
\alpha\cdot \triangledown}), V \}+\\ \qquad 2ic\hbar\epsilon({
\alpha\cdot \triangledown})-2\varepsilon V-2\varepsilon
mc^2\beta+\epsilon^2
\end{array}
\end{equation}
where for the Coulomb potential $V=-\frac{e^2Z}{r}$, $e$ and $Z$
are an electron charge and charge number of the nucleus
respectively. Our further purpose is to solve the eigen problem
for the squared Dirac's equation
\begin{equation}\label{Ham3}
H\Psi=\lambda\Psi
\end{equation}
numerically. It's clear that $\lambda=(E-\epsilon)^2$ where $E$ is
an eigen number for the $h$ operator. In order to reach that goal
one can adapt the usual representation for the eigen function of
the Dirac's equation with spherically symmetric bound potential
\cite{Lab}. Namely let's take
\begin{equation}\label{psi}
\Psi=\left( g \Omega_{jlm}\atop if\Omega_{j\overline lm} \right)
\end{equation}
that is well-known representation for the Dirac's bispinor,
$g=g(r)$ and $f=f(r)$ are upper and lower radial component
functions respectively. For the purpose of numerical
implementation one needs to expand those functions over finite set
of basis functions. We used here finite basis constructed from the
B-splines (see for example \cite{John}). So if one has got the
finite basis set $\{B_i(x)\}_{i=1}^n$ it's straight forward to
gain the generalized symmetric eigen value matrix problem for the
equation (\ref{Ham3})
\begin{equation}\label{matrix}
\left (
\begin{array}{cc}
H^{(1)}& H^{(2)} \\ H^{(3)}& H^{(4)}
\end{array}
\right )x =\lambda
\left (
\begin{array}{cc}
B^{(1)}& 0 \\ 0& B^{(2)}
\end{array}
\right ) x
\end{equation}
Where $H^{(1)}$, $H^{(2)}$, $H^{(3)}$, $H^{(4)}$, $B^{(1)}$,
$B^{(2)}$ are the sub-matrixes $(n\times n)$ of the general
matrixes $H$ and $B$ $(2n\times 2n)$, with the matrix elements
\begin{equation}\label{me1}
B^{(1)}_{ij}=B^{(2)}_{ij}=\left( B_ir^2B_j\right)
\end{equation}

\begin{equation}\label{me2}
\begin{array}{l}
H^{(1)}_{ij}=-\hbar^2c^2\left(
(B_ir^2B''_j)+2(B_irB'_j)-l(l+1)(B_iB_j)\right)+
\\ \qquad
m^2c^4(B_ir^2B_j)+(B_ir^2V^2B_j)+2mc^2(B_ir^2VB_j)-
\\ \qquad
2\epsilon
mc^2(B_ir^2B_j)-2\epsilon(B_ir^2VB_j)+\epsilon^2(B_ir^2B_j)
\end{array}
\end{equation}

\begin{equation}\label{me3}
\begin{array}{l}
H^{(2)}_{ij}=-c\hbar\left(
2(B_ir^2VB'_j)+2(1-\kappa)(B_irVB_j)+(B_ir^2V'B_j)\right )+
\\ \qquad
2\epsilon c\hbar\left( (B_ir^2B'_j)+(1-\kappa)(B_irB_j)\right)
\end{array}
\end{equation}

\begin{equation}\label{me4}
\begin{array}{l}
H^{(3)}_{ij}=c\hbar\left(
2(B_ir^2VB'_j)+2(1+\kappa)(B_irVB_j)+(B_ir^2V'B_j)\right )-
\\ \qquad
2\epsilon c\hbar\left( (B_ir^2B'_j)+(1+\kappa)(B_irB_j)\right)
\end{array}
\end{equation}

\begin{equation}\label{me5}
\begin{array}{l}
H^{(4)}_{ij}=-\hbar^2c^2\left( (B_ir^2B''_j)+2(B_irB'_j)-\overline
l(\overline l+1)(B_iB_j)\right)+
\\ \qquad
m^2c^4(B_ir^2B_j)+(B_ir^2V^2B_j)-2mc^2(B_ir^2VB_j)+
\\ \qquad
2\epsilon
mc^2(B_ir^2B_j)-2\epsilon(B_ir^2VB_j)+\epsilon^2(B_ir^2B_j)
\end{array}
\end{equation}
besides $(f)=\int_0^\infty f(r)dr$.

\section{Spectra quality test, $\epsilon=0$ operator}
To examine the grade of the spectra Bi H-like ion was chosen. The
knot net sequence for the B-spline construction was taken in
accordance with the formula $k(i)=\frac{(b-a)}{n^6}i^6+a$. The
boundary conditions were adapted and $f(a)=f(b)=g(a)=g(b)=0$ in
order to prevent the kinetic energy not to be negative. The
outcome of the calculation (diagonalization of (\ref{Ham3}) with
$\epsilon=0$ ) is presented in Table \ref{firsttable}. In Table
\ref{firsttable} the lowest eigen values are compared not only
with the corresponding eigen values gained from Sommerfeld formula
but also with the eigen values received from the virial relations
(\ref{vr1}, \ref{vr2}). The virial theorem for the operator $h$
and $h^2$ yields formulae

\begin{equation}\label{vr1}
E=mc^2\langle \Psi| \beta|\Psi\rangle
\end{equation}
and
\begin{equation}\label{vr2}
E^2=m^2c^4+mc^2\langle \Psi| V\beta|\Psi\rangle
\end{equation}
respectively. As one can see from Table \ref{firsttable} the
agreement is quite good. In addition to this different spectra
(especially with $\kappa>0$) were analyzed and it was found out
the absence of so called spurious states. To avoid problems with
spurious states special care needs while numerical implementation
of the equation $h\Psi=E\Psi$ \cite{Shab}.

\begin{table}[!h]
\caption[]{Several lowest energy levels for Bi H-like ion with
$l=0$, $j=\frac{1}{2}$. The order of splines $k=6$ and the number
of intervals for the B-splines construction is $n=130$. Parameters
for the knot sequence construction are $a=10^{-15}$ and $b=10$
respectively. $E$ stands for self energy value gained by
diagonalization of the equation (\ref{Ham3}).  $E_{S}$ means
values received from Sommerfeld's formula. The virial theorem
tests are also presented. The $E_{vr1}$ and $E_{vr2}$ columns give
the self energies received with formulae (\ref{vr1}) and
(\ref{vr2}) respectively.}
\begin{center}
\begin{tabular}{|c|c|c|c|}
\hline $E_{S}$ & $E$& $E_{vr1}$& $E_{vr2}$ \\ \hline
   -3836.36956&  -3836.36901&   -3836.36498&  -3836.36583\\
\hline
   -984.921152&  -984.921036&   -984.920525&  -984.920581\\
\hline
   -425.768940&  -425.768909&   -425.768766&  -425.768777\\
\hline
   -234.607639&  -234.607627&   -234.607570&  -234.607573\\
\hline
   -148.000373&  -148.000368&   -148.000340&  -148.000340\\
\hline
   -101.712322&  -101.712319&   -101.712304&  -101.712304\\
\hline
   -74.1444640&  -74.1444625&   -74.1444530&  -74.1444528\\
\hline
   -56.4227560&  -56.4227552&   -56.4227489&  -56.4227487\\
\hline
   -44.3652984&  -44.3652978&   -44.3652936&  -44.3652933\\
\hline
   -35.7940987&  -35.7940983&   -35.7940953&  -35.7940951\\
\hline
   -29.4849330&  -29.4849328&   -29.4849305&  -29.4849304\\
\hline

\end{tabular}
\end{center}
\label{firsttable}
\end{table}

\section{The states from continuous part of spectra}
As one can see the outcome of the diagonalization of the equation
(\ref{Ham3}) with $\epsilon=0$ yields for the eigen values
corresponding to the continuous part of spectra to be doubly
degenerated. In general case the wave function $\Psi$ (for that
part of spectra) is a mix of the couple of functions $\varphi_{+}$
(really the self energy function of the $h$ operator with the
corresponding self energy $E>0$) and $\varphi_{-}$ (really the
self energy function of the $h$ operator with the corresponding
self energy $-E$)
\begin{equation}\label{f}
\Psi=C_1\varphi_{+}+C_2\varphi_{-}
\end{equation}
with some arbitrary coefficients $C_1$ and $C_2$. Utilizing the
properties of the self energy functions (\ref{hfp}), (\ref{hfm})
and (\ref{hf})

\begin{equation}\label{hfp}
h\varphi_{+}=E\varphi_{+}
\end{equation}

\begin{equation}\label{hfm}
h\varphi_{-}=-E\varphi_{-}
\end{equation}

\begin{equation}\label{hf}
h\Psi=C_1E\varphi_{+}-C_2E\varphi_{-}
\end{equation}
one can easy separate the positive and negative spectra functions.
\begin{equation}\label{dif1}
h\Psi+E\Psi=2C_1E\varphi_{+}
\end{equation}

\begin{equation}\label{dif2}
h\Psi-E\Psi=-2C_2E\varphi_{-}
\end{equation}
Finally the formulae (\ref{dif1}) and (\ref{dif2}) present the
desired functions. Everyone can also easy see that those functions
are orthogonal (They really must be!). The formula (\ref{proof})
gives the proof of the fact.

\begin{equation}\label{proof}
\begin{array}{l}
\langle \varphi_{+}| \varphi_{-}\rangle=-(4C_1C_2E^2)^{-1}\langle
h\Psi+E\Psi| h\Psi-E\Psi\rangle=\\
\qquad\qquad\qquad-(4C_1C_2E^2)^{-1}\langle
\Psi|(h+E)(h-E)|\Psi\rangle=
\\ \qquad\qquad\qquad-(4C_1C_2E^2)^{-1}\langle \Psi|h^2-E^2|\Psi\rangle=0
\end{array}
\end{equation}
To conclude this section we can say that there is no problem to
get full spectra and eigen functions of Dirac's one-electron
Hamiltonian from the squared one.

\section{The $\epsilon\neq 0$ operator}
The case of the operator (\ref{Ham3}) with $\epsilon\neq0$ is very
interesting one. When $\epsilon=0$ the spectra of the operator
(\ref{Ham3}) looks like one of the Schrodinger equation, namely we
have two part spectra, the lowest energies are bound states and
energies of the states from continuous spectra above them. The
$\epsilon\neq0$ transforms spectra of Dirac's equation even more,
namely appropriate choice gives possibility to make any energy
level the lowest one. This possibility was tested and gave
positive results. The most important thing is that eigen functions
and values (energies) could be found separately from each other as
the lowest eigen values of the squared Hamiltonian (\ref{Ham3}).

\section{Conclusions and future perspective}
The results of the present work are following. Another way to get
spectra and corresponding eigen functions of one electron Dirac's
equation is presented. It's shown the absence of spurious states.
It was pointed the very attractive method to get highly excited
states of the Hamiltonian separatly each other as the lowest eigen
values of some squared Hamiltonians. The further purpose of author
is a generalization of the method for many electron case.

\end{document}